\documentclass[aps,prl,amssymb,twocolumn,showpacs,supersriptaddress,groupeaddress]{revtex4-1}
\usepackage{amssymb}
\usepackage{amsmath}
\usepackage{graphicx}
\usepackage{bm}
\usepackage{caption}
\usepackage{subcaption}
\usepackage{bm}
\usepackage{natbib,hyperref}
\usepackage{hyperref}
\usepackage{color}
\usepackage{multirow}
\usepackage[titletoc]{appendix}
\hypersetup{colorlinks=true, 
    linkcolor=red,          
    citecolor=magenta,        
    filecolor=gree,      
    urlcolor=blue           
}

\graphicspath{{images/}}

\begin{document}


\title{Electric selective activation of memristive interfaces in TaO$_x$-based devices}

\author{C. Ferreyra $^{1}$, M. J. S\'anchez $^{2}$, M. Aguirre$^{3,4,5}$, C. Acha$^{6}$, S. Bengi\'o$^{2}$, J. Lecourt$^{7}$, U. L\"uders$^{7}$, D. Rubi$^{1}$}
\affiliation{$^{1}$ GIyA and INN-CONICET, CNEA, Av. Gral Paz 1499 (1650),San Mart\'{\i}n, Buenos Aires, Argentina.\\
$^{2}$ INN-CONICET, Centro At{\'{o}}mico Bariloche and Instituto Balseiro, (8400) San Carlos de Bariloche, Argentina.\\
$^{3}$ Departmento de F\'{\i}sica de Materia Condensada, Universidad de Zaragoza, Pedro Cerbuna 12  50009 Zaragoza - Spain.\\ 
$^{4}$ Laboratorio de Microscop\'{\i}as Avanzada (LMA), Instituto de Nanociencia de Arag\'on (INA)-Universidad de Zaragoza, C/Mariano Esquillor s/n. 50018 Zaragoza, Spain.\\
$^{5}$ Instituto de Ciencias de Materiales de Arag\'on (ICMA), Universidad de Zaragoza, Zaragoza, Spain.\\ 
$^{6}$ Depto. de F\'{\i}sica, FCEyN, Universidad de Buenos Aires and IFIBA, UBA-CONICET, Pab I, Ciudad Universitaria, Buenos Aires (1428), Argentina.\\
$^{7}$ CRISMAT, CNRS UMR 6508, ENSICAEN, 6 Boulevard Maréchal Juin,14050 Caen Cedex 4, France.}




\date{\today}%
\begin{abstract}

The development of novel devices for neuromorphic computing and non-traditional logic operations largely relies on the fabrication of well controlled memristive systems with functionalities beyond standard bipolar behavior and digital ON-OFF states. In the present work we demonstrate  for Ta$_2$O$_5$-based devices that it is possible to selectively activate/deactivate two series memristive interfaces in order to obtain  clockwise  or counter-clockwise multilevel squared  remanent resistance loops, just by controlling the  (a)symmetry of the applied stimuli and independently of the nature of the used metallic electrodes. Based on our thorough characterization, analysis and modeling, we show that the physical origin of this electrical behavior relies on controlled oxygen vacancies electromigration between three different zones of the active Ta$_2$O$_{5-x}$ layer: a central -bulk- one and two quasi-symmetric interfaces with reduced TaO$_{2-h(y)}$ layers. Our devices fabrication process is rather simple as it implies the room temperature deposition of only one CMOS compatible oxide -Ta-oxide- and  one  metal, suggesting  that  it  might  be  possible  to take advantage of these properties at low cost and with easy scability. The tunable opposite remanent resistance loops circulations with multiple -analogic- intermediate stable states allows mimicking the adaptable synaptic weight of biological systems and presents potential for non-standard logic devices.

\end{abstract}
\maketitle

\section{Introduction}
\label{int}
Non-volatile resistive switching (RS) devices, usually called memristors, are intensively studied due to their potential as resistive random access memories (ReRAM) \cite{ielmini_2016}, novel logic units \cite{borghetti_2010} and neuromorphic computing \cite{yu_2017} devices. RS has been found for a large number of simple and complex transition metal oxides, including TaO$_x$ \cite{prakash_2013}. Outstanding memristive figures have been reported in this system, including endurances larger than 10$^{12}$ cycles \cite{lee_2011}, ON-OFF ratios up to 10$^6$ and retention times greater than 10 years \cite{shi_2017}. RS in simple oxides such as TaO$_x$ has been reported as bipolar and the physical mechanism was usually attributed to the creation and disruption of conducting oxygen vacancies (OV) nano-filaments \cite{prakash_2013}. This process usually takes place in a region close to a metal electrode with a high work function such as Pt, which in contact with a n-type insulating oxide forms a Schottky barrier \cite{baek_2016}. The drift diffusion of OV to and from the interface under the action of the external electrical field modifies the Schottky barrier and modulates its resistance. The application of negative voltage to the electrode attracts positively charged OV to the interface and triggers the transition from high resistance (HR) to low resistance (LR), usually called SET process. The opposite transition is achieved with positive voltage and is usually called RESET process. In this way, the circulation of the current-voltage (I-V) curve is clockwise in the positive current-positive voltage quadrant, while the remanent resistance vs. voltage curve (usually called Hysteresis Switching Loop, HSL \cite{rozenberg_2010}) displays a counter-clockwise behavior.

The possibility of opposite (counter-clockwise) I-V and (clockwise) HSL circulations was also shown for TiO$_2$/Pt interfaces, and attributed to oxygen exchange between Pt and TiO$_2$ at the local position of the filament \cite{zhang_2018}. Opposite switching polarities were also reported for TaO$_x$ devices and attributed to a competition between ionic motion and electronic - i.e charge trapping- effects \cite{latorre_2017}, to the formation of filaments with either conical or hour-glass shapes \cite{park_2015,park_2017} or to volumetric oxygen vacancies exchange between TaO$_x$ layers with different stochiometries \cite{yang_2012}. The coexistance of two bipolar memristive regimes with opposite polarities in a single device allows to increase or decrease the resistance of the device with stimulus of different amplitudes but the same polarity, which has potential for the development of beyond von Neumann novel computing devices \cite{zhang_2018}. 

Asymmetric devices (i.e. an insulating oxide sandwiched between two electrodes of different metals) with an active interface and an ohmic non-active one were reported to display squared HSL \cite{ghenzi_2013,ghenzi_2014}. On the contrary, symmetric systems  presenting two similar metal/oxide memristive active interfaces in series were reported to display HSLs with the so-called ``table with legs'' (TWL) shape \cite{chen_2005,rozenberg_2010}. In these cases, the two interfaces behave in a complementary way: when one switches from low resistance (LR) to high resistance (HR), the other switches inversely. Symmetric Pt/TiO$_2$/Pt devices stimulated with symmetric stimuli displayed complex I-V curves with multiple transitions, which was interpreted as bipolar switching taking place simultaneously at both Pt/TiO$_2$ interfaces \cite{jeong_2009}. 

In  the  present  paper  we  show  for  Ta$_2$O$_5$-based  devices  that  a  careful  tuning  of  the  stimuli  protocol  allows  to  selectively  activate/deactivate  the  contribution of two memristive interfaces to the overall RS behavior,
allowing to obtain both clockwise/counter-clockwise multilevel squared HSL or a ``table with legs'' behavior, where both interfaces are active simultaneously.

We show that the (a)symmetry of the electrical response can be controlled by appropriately tuning the electroforming process and the subsequent excitation protocol, independently of the electrodes symmetry. From a combination of I-V curves analysis and modeling, togheter with numerical simulations based on the Voltage Enhanced Oxygen Vacancy migration (VEOV) model \cite{rozenberg_2010}, adapted for binary oxides-based devices  \cite{ghenzi_2013}, we achieve a thorough understanding of the observed phenomenology, which is rationalized in terms of oxygen vacancies electromigration between a central (bulk) zone of the active Ta$_2$O$_5$ layer and its interfaces with metallic TaO$_x$ (x $<$ 2). Finally, the potential technological implications of the observed tunable opposite HSL circulations with multiple intermediate stable states are discussed.

\section{Results}
\label{resu}

\begin{figure}[!htb]
\includegraphics[scale = 0.5]{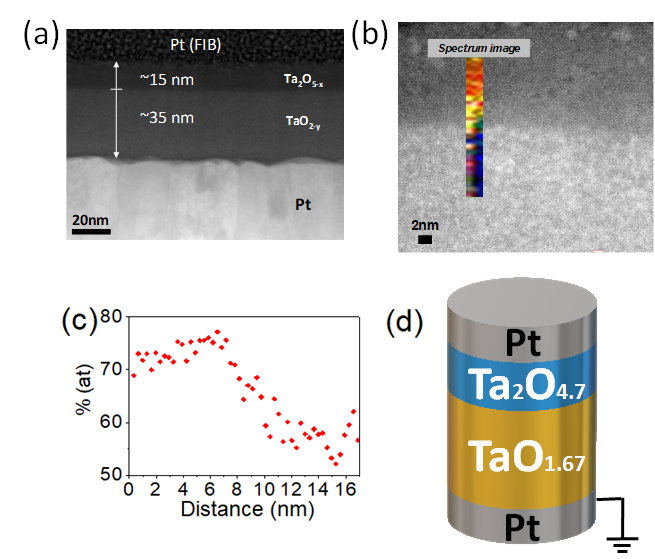}
\caption{(a)  STEM-HAADF cross-section corresponding to a TaO$_x$ bilayer grown at 0.1 (15nm, layer A) and 0.01 (35nm, layer B) mbar of O$_2$; (b) STEEM-HAADF-EELS line scan (colour map) of oxygen concentration at a zone close to the interface between both layers. The line scan starts in the more oxidized layer and ends in the more reduced one. See text for details; (c) Oxygen concentration quantification from data presented in (c); (d) Sketch showing the geometry and average chemical composition of the pristine devices. The bottom Pt electrode was grounded and electrical stimuli was applied to the top one.
}
\label{f1}
\end{figure}

We have fabricated and characterized Ta-oxide  bilayers grown on platinized silicon at different oxygen pressures. We will first describe the performed transmission electron microscopy characterization. Figure \ref{f1}(a) shows a STEM-HAADF cross section, evidencing that the thicknesses of the Ta-oxide layers grown at 0.01 (layer B) and 0.1 (layer A) mbar of O$_2$ are 35nm and 15nm, respectively. Both layers are amorphous and differ in their contrast. A brighter image is found for the B layer, indicating a higher relative concentration of Ta (see the blow up displayed in Figure S1 of the Supp. Information), consistently with a more reduced oxide and the lower oxygen pressure used for its growth. 
Figure \ref{f1}(b) shows an STEEM-HAADF-EELS map of the bilayer indicating a higher relative concentration of oxygen at the top A layer. Performed quantifications, as the one shown in Figure\ref{f1} (c), indicate average oxygen concentrations of $\approx 70$ \%at and $\approx$ 60\%at for A and B layers, respectively. This suggests average stochiometries for A and B layers of Ta$_2$O$_{4.70}$  and TaO$_{1.67}$, indicating in the latter case a mixture of $\approx$ 67 \% TaO$_2$ and $\approx$ 33 \% TaO. Consistent results are obtained from STEEM-HAADF-EDX quantification, as shown in Figure S1 of the Supp. Information \cite{si}. 

\begin{figure}[!htb]
\includegraphics[scale = 0.55]{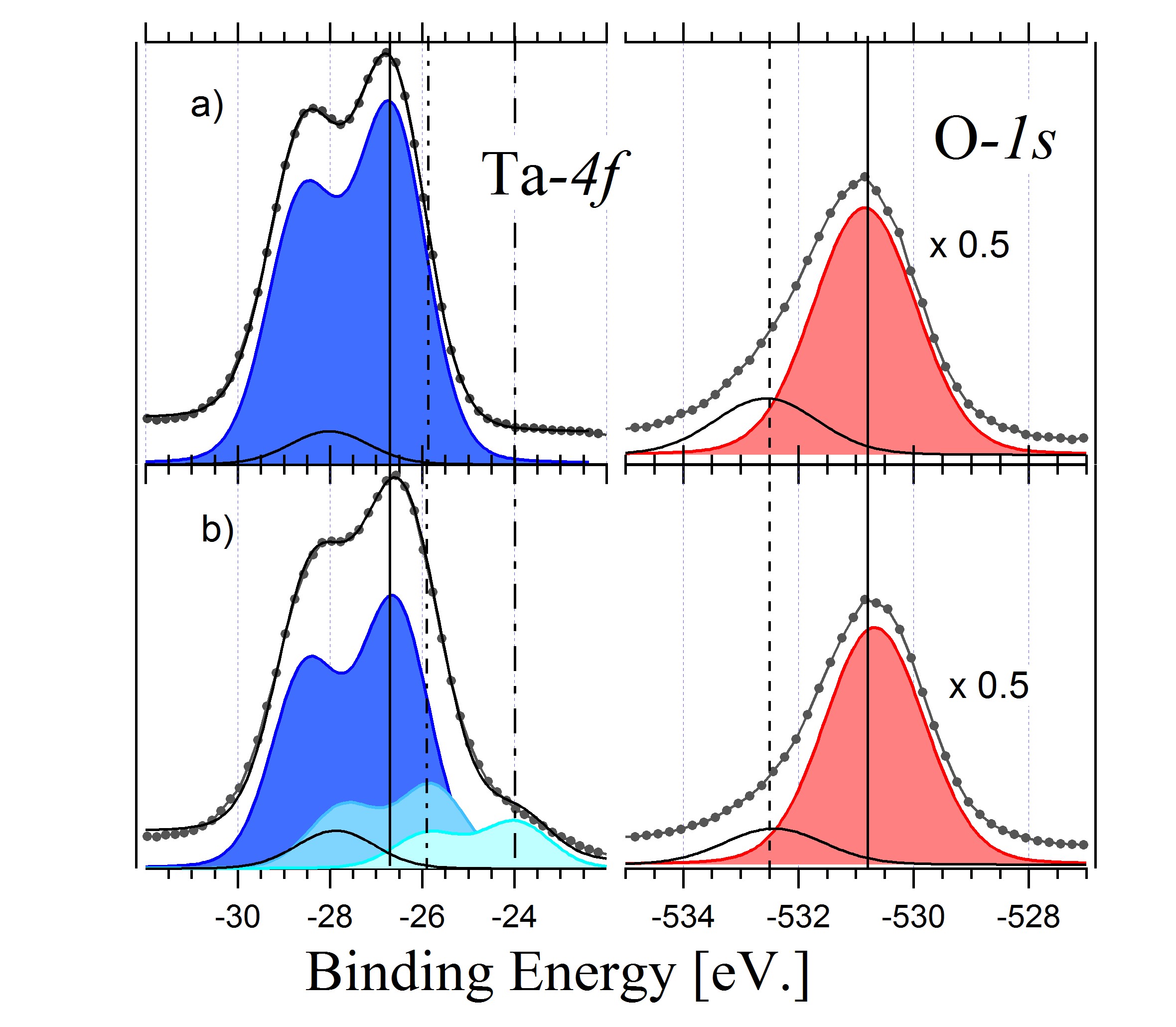}
\caption{(a)  X-ray photoemission Ta-$4f$ and O-$1s$ spectra recorded on a TaO$_x$ film grown at 0.1 mbar of O$_2$; (b) Same spectra for a TaO$_x$ film grown at 0.01mbar of O$_2$. The references for Ta$_2$O$_5$ at 26.7 eV, TaO$_2$ at 25.9 eV, TaO at 24 eV, lattice oxygen at 530.8 eV and adsorbed O$_2$ at 532.5 eV are shown as vertical lines. The areas under each component of the spectra are displayed with different colours. See text for details.}
\label{f2}
\end{figure}

XPS measurements were performed to obtain information of the films surface chemical composition and Ta oxidation states. Both Ta-$4f$ and O-$1s$ spectra are shown in Figure \ref{f2} (a) and (b) for TaO$_x$ films grown at 0.1 mbar and 0.01 mbar of O$_2$, respectively. The spectra were fitted using a Voight function for each peak plus a Shirley-type background. The total fitted intensities along with the experimental ones are shown in each spectrum. Ta-4f spectra components are characerized by two identical peaks corresponding to the spin-orbit split between 4f$_{7/2}$ and 4f$_{5/2}$ levels, with relative intensities of 3:4. The Ta-4f spectrum of the film grown at 0.1mbar of O$_2$ display one component corresponding to Ta$_2$O$_5$ and for the film grown at 0.01mbar of O$_2$ three components were identified: a major one at binding energy 26.7 eV ascribed to Ta$_2$O$_5$ and two minor ones at binding energies 25.9 eV and 24eV, ascribed to TaO$_2$ and TaO respectively. Besides, there is a O-2s component at 27.8eV. The presence of some Ta$_2$O$_5$ at the spectra of the film grown at low O$_2$ pressure, absent in the case of layer B (buried) in the bilayer analyzed by STEM-HAADF, is related to the ex-situ character of the XPS experiments and indicates surface re-oxidation upon exposure of the film to ambient pressure. It is worth noting that no sputtering process was performed on the films, as it is a possible source of vacancies creation and formation of metastable suboxides \cite{sipmson_2017}. The O$-1s$ spectra display a major component at 530.8 eV, ascribed to the metal oxide and a minor component at 532.5 eV, attributed to adsorbed molecules.

Based on combined STEM-HAADF and XPS analysis, we characterize our virgin devices as Pt/Ta$_2$O$_{4.70}$/TaO$_{1.67}$/Pt, as shown in the sketch displayed on Figure \ref{f1}(d).    

We describe now the electrical characterization. The bottom Pt electrode was grounded and the electrical stimulus (voltage) was applied to the top Pt electrode.

\begin{figure}[!htb]
\includegraphics[scale = 0.29]{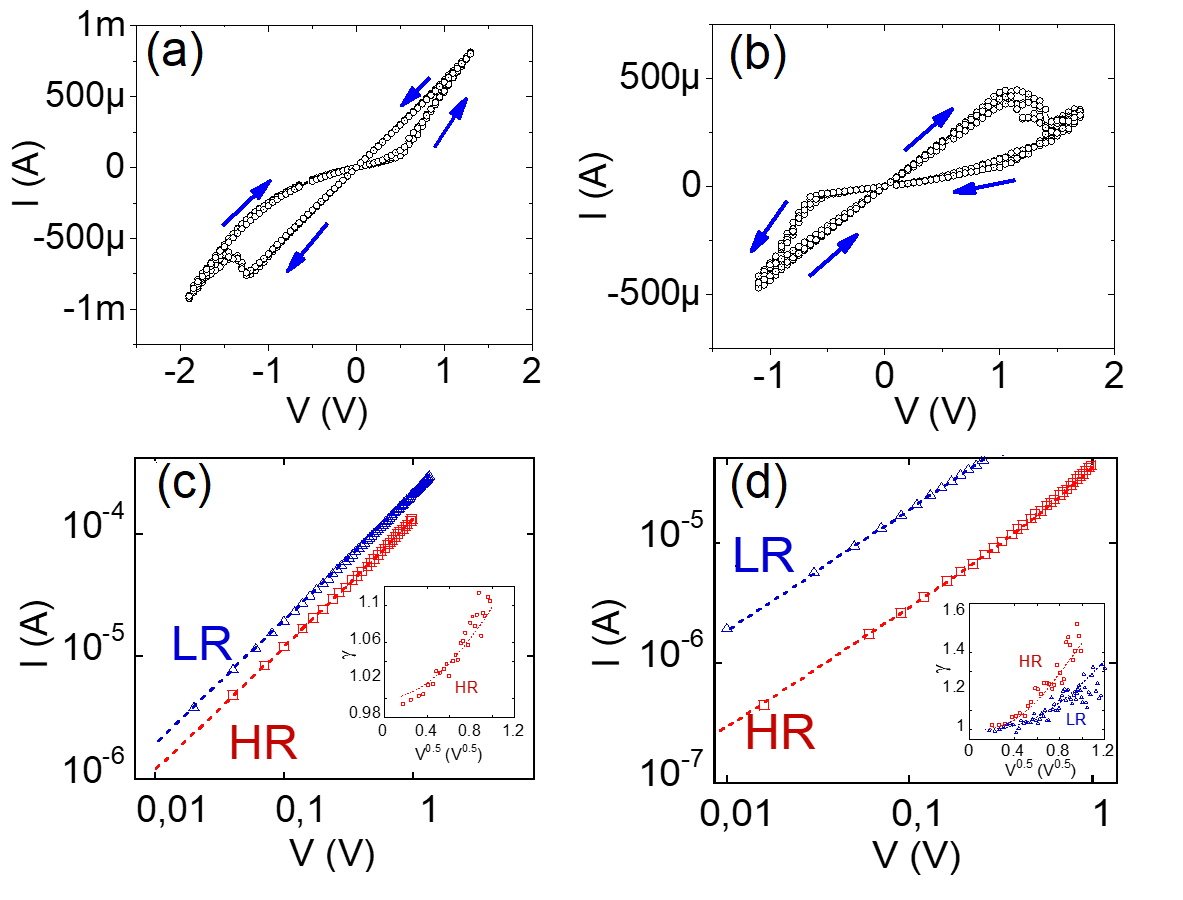}
\caption{I-V response of a device with 28x10$^3$ $\mu$m$^2$ area, obtained for voltage excursions between  $V_{max}$ = +1.5 V and $V_{min}$ = -2 V (a), and  for  $V_{max}=$2 V and $V_{min}$=-1.5 V (b), where the circulation of the I-V curve is inverted. (c),(d) Fittings -in dashed lines- performed on I-V curves recorded for a 6.4x10$^3$ $\mu$m$^2$ device, for the high and low resistance states, labeled respectively HR and LR. The used equivalent circuit is described in the text and Supp. Information. The insets show the  associated $\gamma$ vs V$^{1/2}$ representation for  the LR ans HR.  For better visualization, the LR state was not included in the inset of panel (c). The dashed line corresponds to the theoretical $\gamma$ curve, extracted from the I-V curves fittings.}
\label{f3}
\end{figure}

The devices were initially in a low resistance state (100 $\Omega)$; an electroforming process with negative stimulus (i.e. -3 V pulses, with a 1 ms width) was necessary to increase the devices resistance to the range of k$\Omega$ and activate their memristive behaviour. After forming, dynamic current-voltage (I-V) curves were obtained by applying a sequence of voltage pulses of different amplitudes (0$\rightarrow V_{max} \rightarrow -V_{min} \rightarrow 0$, with a time-width of 1ms and a step of 100 mV), with the current measured during the application of the pulse. Additionally, after each voltage pulse a small reading voltage was applied to allow the current to be measured and the remnant resistance state determined, obtaining the  HSL. We have found that the devices electrical response is highly dependant on both  $V_{max}$ and $V_{min}$.

\begin{figure}[!htb]
\includegraphics[scale = 0.4]{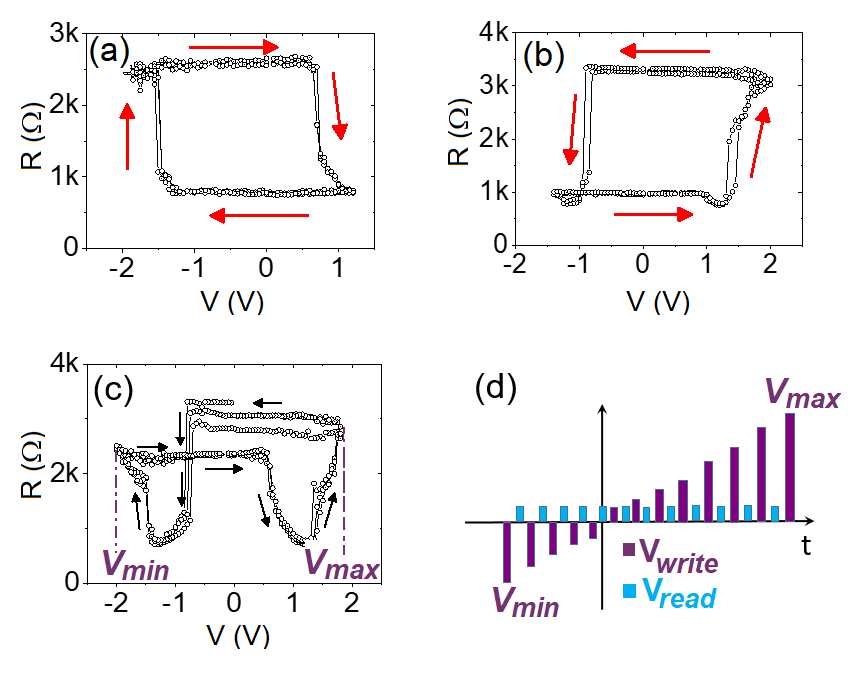}
\caption{HSL's with (a) CW circulation and (b) CCW circulation, for asymmetric voltage excursions; (c) TWL HSL obtained for symmetric voltage excursions, showing two complementary active interfaces. The same device than in Figures \ref{f3} (a) and (b) was used ; (d) Voltage protocol used to obtain dynamic I-V curves and HSL. $V_{max}$ ($V_{min}$) is the maximum (minimum) voltage excursion for the write pulses. Notice the different values of $V_{min}$ and $V_{max}$ for the CW and CCW HSL's of panels (a) y (b), respectively.}
\label{f4}
\end{figure}

Figures \ref{f3} (a) and \ref{f4} (a) display respectively, the dynamic I-V curve and HSL obtained for a device with 28x10$^3$ $\mu$m$^2$ area, for $V_{max}$ = +1.5 V and $V_{min}$ = -2 V. It is found that the transition from HR$_2$ to LR$_2$ (SET process, the notation of the resistive states is chosen to be consistent with the fittings and simulations to be described below) is achieved with positive stimulus ($V_{S}$ $\approx$ +1V) while the opposite transition (RESET) is obtained with negative voltage ($V_{R}$ $\approx $ -1.2 V). The LR$_2$ and HR$_2$ states are 0.8 k$\Omega$ and 2.5 k$\Omega$  respectively, giving an ON-OFF ratio of 3.1. The stability and reproducibility of the curves is remarkable, as shown in Figure S2 of the Suppl. Information for 200 consecutive cycles. In addition, retention times up to 10$^4$ s were checked for both resistive states. 

From these measurements, it is established that the HSL present a clockwise (CW) circulation. Remarkably, this circulation is inverted to a counter-clockwise (CCW) behaviour when the maximum voltage excursion are inverted to $V_{max}$ = +2 V and $V_{min}$ = -1.5 V, as shown in Figures \ref{f3} (b) and \ref{f4} (d), respectively. In this case, the transition from HR$_1$ to LR$_1$ (SET process) is achieved with negative stimulus ($V_{S}$ $\approx $ -1V) while the opposite transition (RESET) is obtained with positive voltage ($V_{R}$ $\approx $ +1.5 V). The LR$_1$ and HR$_1$ states in this case are 1  k$\Omega$ and  3.3k$\Omega$  respectively, giving an ON-OFF ratio of 3.3. Again, an excellent stability upon consecutive cycling is found (200 cycles), as shown in Fig. S2 of the Supp. Information, and retention times for up 10$^4$s were also checked. These results clearly show that the circulation of the HSL can be controlled and tuned on the same device by properly choosing the voltage excursions of applied stimuli. The inverse dependance of LR$_1$ and LR$_2$ states with the device area (shown in Figure S3 of the Supp. Information) indicates the existance of conducting paths comprising the complete device area.  

Interestingly, if the voltage excursions are enlarged and symmetrized to $V_{max} =-V_{min}$=2 V, the  HSL change from squared to a TWL-like shape, as shown in Figure \ref{f4} (c). The squared HSLs of Figures \ref{f4} (a) and (b) correspond to a single active interface while the TWL of Figure \ref{f4} (c) corresponds to two complementary active interfaces; that is, when one switches from low to high resistance the other one changes inversely \cite{rozenberg_2010}. These results clearly indicate that our two memristive series interfaces can be selectively decoupled and activated by means of the application of proper electrical stimuli, and goes against previous claims about the ``simultaneous" memristive behavior of two series interfaces \cite{jeong_2009}.

\begin{figure}[!htb]
\includegraphics[scale = 0.45]{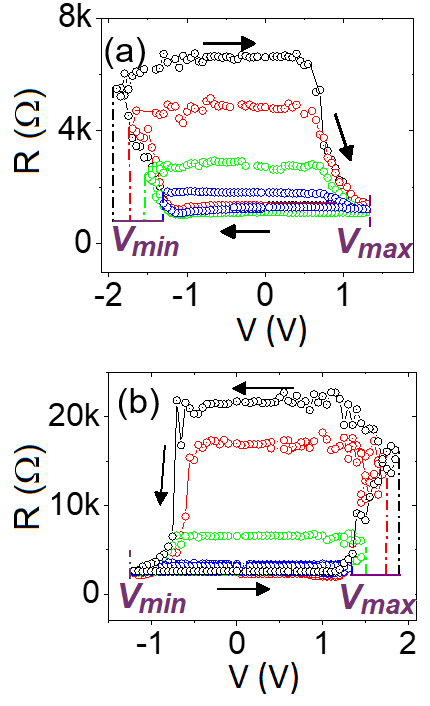}
\caption{Multilevel HSL recorded on a device with 11.3x10$^3$ $\mu$m$^2$ area for (a) CW circulation and (b) CCW circulation. In (a) V$_{max}$ is fixed while V$_{min}$ is changed; in (b) V$_{min}$ is fixed while V$_{max}$ is changed.}
\label{f5}
\end{figure}

We have found that for both (squared) HSL circulations multilevel states are possible. Figure \ref{f5} (a) shows, for a device with 11.3x10$^3$ $\mu$m$^2$ area, that different non-volatile resistance levels can be achieved for CW HSLs upon increasing $V_{S}$ from 1.3 to 1.9 V, while keeping $V_{R}$ fixed at -1.9 V. In a similar way, Figure \ref{f5} (b) shows a multi-level behavior for a CCW HSL upon changing $V_{S}$ from -1.3 V to -1.9 V while keeping $V_{R}$=1.2 V fixed. 
We note that this device presents higher remanent resistance states due to its lower area in relation to the previously described devices of Figures \ref{f3} and \ref{f4}. 

The multilevel resistance states show that our devices behave in an analogic way, which is a key feature to mimic the adaptable sinaptic weight of biological synapsis.

\begin{figure}[!htb]
\includegraphics[scale=0.40]{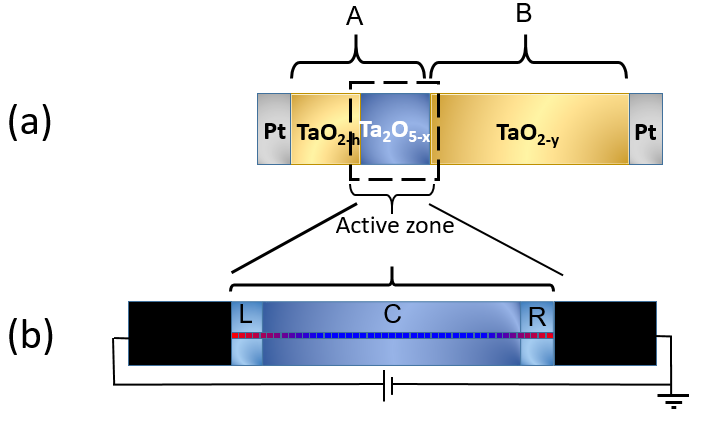} \hfill
\caption{(a) Schematic diagram of the post-formed device. The initial asymmetric geometry turns into   a quasi-symmetric Pt/TaO$_{2-h}$/Ta$_2$O$_{5-x}$/TaO$_{2-y}$/Pt stack after forming; (b) Sketch of the post-forming active region for the memristive behavior and scheme of the 1D chain of nanodomains assumed for the numerical simulations with the VEOV model. See text for details.}
\label{f6}
\end{figure}

In order to understand the described behavior, we start with the forming process. Virgin devices are  found in a low resistance state, indicating that the more oxidated A layer (with an average stochiometry Ta$_2$O$_{4.70}$, according to STEEM-HAADF experiments) presents a substantial higher conductivity that the one expected for the insulating stochiometric compound. This can be explained both by the presence of strong oxygen deficiency and to some material inhomogeneity at the nanoscale (see Figure S1 of the Supp. Information), which allow the presence of percolative low resistance Ta-rich paths bridging the nearby layers. However, as both TaO$_2$ and TaO present at the B layer are metallic \cite{prakash_2013}, the A layer still dominates the resistance of the oxide bilayer. The application of negative voltage to the top (left in Figure \ref{f6} (a)) electrode triggers an electroforming process that increases the resistance of the device by inducing the migration of OVs within the A layer, as the electrical field acting on this layer is higher that the one acting on the more reduced B layer. Oxygen ions are pushed down within the A layer (from left to right in the scheme of Figure \ref{f6} (a)), leading to the formation of a nearly stochiometric Ta$_2$O$_{5-x}$, resistive, layer and a strongly reduced, metallic, layer in contact with the Pt top (left zone in Figure \ref{f6} (a)) electrode. The presence of this reduced layer is infered from the symmetric and non-rectifying behavior observed in the post-forming I-V curves (Figures  \ref{f3} (a) and (b)) which rule out the presence of Ta$_2$O$_{5-x}$ in contact with the left Pt electrode, as the Pt/Ta$_2$O$_{5-x}$ interface is known to present rectifying (Schottky) behavior \cite{zhuo_2013, lee_2011}. 

Interestingly, we have found that if the forming process is done with positive voltage, a rectifying behavior is observed in the I-V curves, indicating that in that case the formed Ta$_2$O$_{5-x}$ layer is in contact with the top Pt electrode, consistently with the inversion of the forming polarity (see Figure S4 of the Supp. Information). We  return now to the negative forming scenario.

Unlike the case of Ti-oxides, which display a large number of stable sub-stochiometric phases with less oxygen content than TiO$_2$ \cite{okamoto_2001}, the commonly accepted Ta-oxide phase diagram only allows stable Ta$_2$O$_5$ and Ta(O) -that is metallic Ta with some diluted oxygen- phases \cite{garg_1996}. TaO$_2$ has been considered as a metastable phase \cite{yang_2010}, but recent reports claim that it might be stable with an even lower potencial energy than Ta$_2$O$_5$ \cite{wei_2008,yang_2010b}. This suggests that the as-grown A layer, with nominal Ta$_2$O$_{4.70}$ composition, might be a metastable phase that is phase separated during forming, due to Joule heating \cite{yang_2010}, into nearly stochiometric stable Ta$_2$O$_{5-x}$ and TaO$_{2-h}$ layers (x,h $<<$ 1), as depicted in Figure \ref{f6} (b). If we assume that the forming process only involves oxygen redistribution within the A layer, it can be estimated that the thickness of Ta$_2$O$_{5-x}$ and TaO$_{2-h}$ layers layers are about 90 \%  and 10 \%  of the initial A layer thickness. 

We remark that in the post forming  state the  Ta$_2$O$_{5-x}$ resistive layer is sandwiched between two reduced metallic TaO$_{2-h}$ and TaO$_{2-y}$ layers, giving a quasi-symmetric geometry, as shown in the zoomed sketch of Figure \ref{f6} (b), which explains the HSL with TWL-like  shape obtained for symmetric stimuli 
(Figure \ref{f4} (c)), typical of symmetric devices.

The scenario proposed above is supported by the following analysis and fittings performed on the post forming I-V curves, considering
the $\gamma$ = dLn(I)/dLn(V) parameter representation \cite{Acha_2017}. This method
proved to be useful in order to reveal the presence of a mixture of conduction mechanisms~\cite{Acha_2016,Acevedo_2017}, as usually found in oxide-based devices \cite{Acha_2011,Cerchez_2013,BLASCO_2015}. As can be observed
in the insets of Figures~\ref{f3} (c) and (d),
associated with both CCW and CW HSLs of devices with 6.4x10$^3$ $\mu$m$^2$ area, the dependence
of $\gamma$ vs V$^{1/2}$ indicates the existence of an ohmic
conduction ($\gamma \simeq$ 1 at low voltages) in parallel with a
mild non-linear space charge limited current (SCLC) conduction ($\gamma$ increases smoothly with V and remains $<$2).

The simplest schematic circuit representation derived from the
voltage dependence of the $\gamma$ parameter consists in two parallel SCLC-ohmic conducting channels in
series. The experimental I-V curves were fitted by considering the equivalent circuit and related equations included in the Supp. Information. The performed fits give an excellent reproduction of the experimental I-V characteristics as well as the associated $\gamma$ vs. V$^{1/2}$ curves, as shown in  Figures ~\ref{f3} (c) and (d). We notice that no good fitting was obtained if Schottky emission is assumed for the non-linear element of the equivalent circuit. This fact, together with the already discussed non-rectifying behavior observed in the I-V curves (Figures  \ref{f3} (a) and (b)) rule out the presence of Ta$_2$O$_{5-x}$ in contact with the left Pt electrode, giving consistency to the proposed post-forming scenario.

\begin{table*}[t]
\caption{Values in k$\Omega$ of the equivalent-circuit elements obtained by
fitting the experimental I-V characteristics with the corresponding equations (see Supp. Information). Values at 0.1 V and 1 V are indicated for the non-linear SCLC element.}
    \begin{tabular}{|c|c|c|c|c|c|c|c|c|}
        \hline  \hline
        \textbf{HSL}      & \textbf{STATE} & \textbf{R$^{eq}_{1}$} & \textbf{R$^{NL}_{1}$(0.1V)} & \textbf{R$^{NL}_{1}$(1V)} & \textbf{R$^{eq}_{2}$}  & \textbf{R$^{NL}_{2}$(0.1V)} & \textbf{R$^{NL}_{2}$(1V)} & \textbf{R$_{rem}$} \\
        \hline \hline
  \multirow{2}{*} \bf{CCW}    &        HR     &    34.2    &    1700   &    70  &  $\leq$0.8     &  - &  - & 34.3  \\ %
                                    &        LR     &    5.2     &    620  &    25   &  $\leq$0.1     &  -  &  - & 5.3\\ %
                                    \hline
       \multirow{2}{*} \bf{CW}      &        HR     &    $\leq$0.1     &    -   &  -   &  8.4     &  770  &    8 & 8.4  \\ %
                                    &        LR     &    $\leq$0.1     &    -   &  -   &  5.1     &  500  &    50  & 5.1 \\ %

        \hline \hline
    \end{tabular}

    \label{tab:ajustes}

\end{table*}

The memristive and non-linear behaviours are nicely reflected in the extracted values
of the equivalent circuit elements, listed in Table~\ref{tab:ajustes}. As can be observed, the electric response associated with the CCW HSL is dominated by R$_1$ and R$^{NL}_{1}$ elements, as the R$_2$ resistor is highly metallized, shortcircuiting the R$^{NL}_{2}$ non-linear element. The remnant resistance measured al low-voltage bias (0.1 V) is essentially the equivalent resistance R$^{eq}_{1}$ between R$_1$ and R$^{NL}_{1}$. When the electric response switches to the CW HSL, the situation is reversed:  R$_{1}$ shortcircuits the non-linear R$^{NL}_{1}$ element and the resistance of the device is dominated by the parallel combination between R$_{2}$ and the non-linear R$^{NL}_{2}$ element. As can be seen in Table~\ref{tab:ajustes}, the determined equivalent resistances at low voltage (100mV) perfectly match the measured remanent values R$_{rem}$.

The physical origin of the observed memristive behavior can be understood in terms of OVs electromigration  between a (bulk) central zone,  the Ta$_2$O$_{5-x}$ layer, and the interfacial zones close to TaO$_{2-h}$ and Ta$_{2-y}$ layers, respectively.
To achieve a behavior consistent with the migration of OVs induced by the electric
field, the CCW HSL should be represented by an active
SCLC-ohmic zone, represented by the parallel R$_1$ and R$^{NL}_{1}$ circuit elements, which includes the left interface with TaO$_{2-h}$ plus the central Ta$_2$O$_{5-x}$ zone (we recall that SCLC is a bulk conduction mechanism) where the memristive effects esentially modify this sector, while the ohmic element R$_2$ represents the right interface with TaO$_{2-y}$, characterized by a rather low resistance. On the contrary, for the CW HSL,
the increase of the amplitude of the negative applied pulses
switches the active element to a zone comprising the right interface with TaO$_{2-y}$ plus the central bulk zone, both represented by R$_{2}$ and R$^{NL}_{2}$, increasing its
resistance and concomitantly decreasing the resistance of the R$_1$ ohmic element, which now short-circuits the SCLC R$^{NL}_{1}$ element (see Supp. Information for a sketch depicting these situations).

In the next section we will carefully address by means of simulations the OV dynamics linked to this behavior.

\section {Numerical Simulations}
\label{model}

Based on the previous description, we analize here the electrical response obtained with the voltage enhanced oxygen vacancies drift (VEOV) model, which allows simulating OV dynamics at the nanoscale, linked to the memristive behavior of our devices.
We start by reviewing  the main equations, and we refer the readers to References \onlinecite{rozenberg_2010, ghenzi_2013},
 for further details.

We consider that the active memristive region is the Ta$_2$O$_{5-x}$ layer, which is assumed to consist of three zones: a central (bulk) one labelled as C and two zones, left (L) and right (R), localized  close to the interfaces with the metallic TaO$_{2-h(y)}$ layers, which present different properties due to the presence of interfacial disorder or defects. A sketch of the assumed memristive active zone is shown in Figure \ref{f6} (b). For the modelling we suppose a  1D chain of  $N= NL+NC+NR$ total sites. The first  $NL$ sites correspond to the  L layer, the following $NC$ sites are assigned to  the central C layer and  the last $NR$ sites are linked to the  R layer and we assume $NC > NL=NR$.
The sites, characterized by their resisitivity,  physically represent  small domains of (sub)nanoscopic dimensions with an initial OVs concentration that  we assume correspond to the post forming  state.

An universal feature of oxides  is that their resistivity is  affected by the precise oxygen stoichiometry. In particular, Ta$_2$O$_5$ behaves as an n-type semiconductor in which oxygen vacancies reduce its resistivity.
Therefore, we write the resistivity $\rho_{i}$ of  each site $i$ as a linear decrescent (most simple) function of the local OV density $\delta_{i}$, namely:
\begin{equation}  \label{e1}
 \rho_{i} = \rho_{0} (1 - A_{i} \delta_{i}),
\end{equation}
where ${\rho_{0}}$ is  the  residual resistivity  for negligible OV concentration ($\delta_{i}=0$) and A$_i$ is a factor that changes between C, L and R layers, satisfying $ A_{i} \delta_{i} < 1 \; \forall i$.  We consider A$_L$, A$_R$ $<$ A$_C$, which implies that the resistivity of the interfacial zones is less sensitive to the presence of OV due to the presence of disorder or defects. Also, the coefficients  $A_{i}$ can be taken for each layer either   smoothly  dependent on the site position or as constants (as we do for simplicity), without affecting the qualitative behaviour of the simulated  results \cite{si}.

Following Equation(\ref{e1}), the total resistivity of the system is given by:

\begin{equation}\label{tr}
\rho \equiv \rho_{s} - \rho_0 \lbrace\sum_{i=1}^{NL} A_{L} \delta_{i} -\sum_{i=NL+ 1}^{N-NR} A_{C} \delta_{i}-\sum_{i=N-NR+1}^{N} A_{R} \delta_{i}\rbrace,
\end{equation}
being $\rho_{s}\equiv N {\rho_{0}}$.

Given an external  voltage $V(t)$ applied  at time $t$, the OV density at site \textit{i}  is updated for each simulation step according to the  rate probability  $p_{ij} = \delta_i (1-\delta_j) \exp(-V_{\alpha} + \Delta V_{i})$,  for a transfer 
from site \textit{i} to a nearest neighbor \textit{j}= \textit{i} $ \pm 1$. Notice that $p_{ij}$  is proportional to the  OV density present at site \textit{i}, and to the available concentration at the neighbour site \textit{j}. In order to restrict the dynamics of OVs to the active region,  we take  $p_{01} =p_{10}=p_{N N+1}=p_{N+1 N}=0$. In addition as the total density of vacancies is conserved, for each simulation step it is satisfied  that $\sum_{i=1}^{N} \delta_{i}=  N \delta_0$, being $\delta_0$ the OV density per site for an uniform  distribution (assumed as known, see Supp. Information section).

In the  Arrhenius factor, $\exp(-V_{\alpha} + \Delta V_{i})$, $\Delta V_i$ is the 
local potential drop  at site \textit{i} defined as ${\Delta V}_i (t) = V_{i}(t) - V_{i-1}(t)$ with $V_i(t) =  V(t) \rho_{i} / \rho$ and $V_\alpha$ the activation energy for vacancy diffusion in the absence of external stimulus. We consider values of  $V_\alpha = V_{L}, V_{C}$ and $V_R$ for the L, C and R layers respectively. All the energy scales are taken in units of the thermal energy $k_{B}T$ \cite{si}.

According to  standard RS experiments, we chose the  stimulus  $V(t)$ as  a linear ramp  following the cycle $0 \rightarrow V_{m1} \rightarrow -V_{m2} \rightarrow 0$.
At each simulation time step $t_k$ we compute the local voltage profile $V_i(t_k)$ and the local voltage drops ${\Delta V}_i (t_k)$ and   employing the probability rates $p_{ij}$ we obtain the transfers between nearest neighboring sites. 
Afterwards the values $\delta_i(t_k)$ are updated to a new set of densities $\delta_i(t_{k+1})$,
with which we compute, at time $t_{k+1}$, the local resistivities  $\rho_i(t_{k+1})$, 
the local voltage drops under the applied voltage $V(t_{k+1})$, and  finally from Eq.(\ref{tr}) the total resistivity $\rho(t_{k+1})$,  to start the next simulation step at $t_{k+1}$. 
Notice that as de VEOV is a 1D model, the conversion from resistivity to resistance is a trivial  scale factor. We refer  to the Supp. Information \cite{si} for further details on  the numerical values of the parameters employed in  the simulations.

\begin{figure}[!htb]
\includegraphics[scale = 0.40]{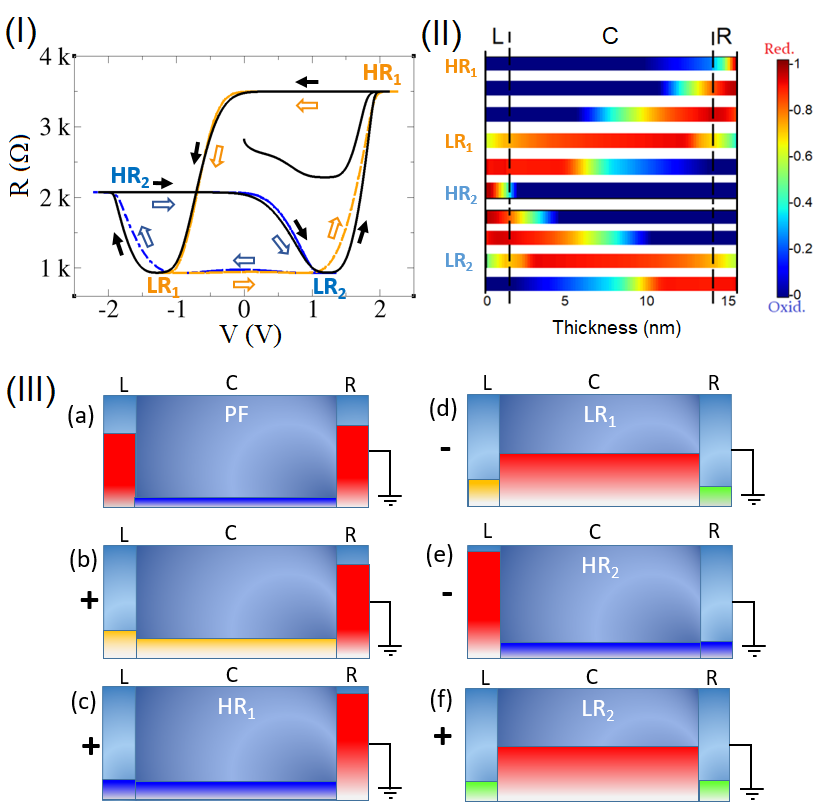}
\caption{I) Simulated HSL's. The TWL-like shape is obtained for symmetric stimuli (black arrows indicate the circulation) and  squared CW and CCW HSLs for asymmetric ones (blue and orange arrows indicate the circulation in the respective HSLs). The  HR and LR states  are  labeled as described in the text; II) Colormap of the OVs density per site (total density normalized to 1) for different resistance states, as labeled respectively in the HSL of panel I);  III) Scheme of the  L, C and  R regions defined in Figure \ref{f6} (b), where the colors  qualitatively show the  different total number of OVs
 in each region for: (a) post forming (PF) state, (c)-(f) HR$_1$, LR$_1$, HR$_2$ and LR$_2$ states, as labeled in panel I).}
\label{f7}
\end{figure}

As it was already described in Sec.\ref{resu}, the  negative forming 
sets the device in a high resistance  state, associated to which we define an initial  OV density profile,  $\delta_i(t_{0}) \; \forall i=1..N$, to start with  the numerical implementation of the VEOV model. This initial OVs profile is chosen to guarantee the post forming  high resistance state, in which  the C zone contributes with the  dominant resistance, while L and R layers present both a lower resistivity due to a large density  of OVs,  and  contribute little to the total resistance (see Figure\ref{f7} III) (a) for a scketch of the initial post forming state).

Figure \ref{f7} panel I)  shows the  TWL-like HSL  obtained with the VEOV model simulations, for a symmetric voltage protocol $V(t)$ following the cycle 0 $\rightarrow$ V$_{m1}$=2.1 V  $\rightarrow $ -V$_{m2}$ =2.1 V $\rightarrow 0$. We start from an initial OVs  distribution compatible with the  post forming state,  scketched in panel III) (a). This initial  profile gives a  resistance of 3K$\Omega$, in perfect agreement  with the  reported experimental value.
The positive stimulus  moves OVs from the L layer into the C layer (see panel III)(b)), initially tending to reduce the resistance. However,  as the applied voltage is  increased a strong electric field develops, moving  OVs further away  to  accumulate finally in the R layer. This gives the
RESET transition to the  HR$_1$ state displayed in the HSL of panel I). The associated OVs density profile along the active region is shown in the colour map of panel II) whilst the total density of OVs in the L, C and R regions respectively,  is sketched in  panel III) (c). 
The HR$_1$ state is mantained  in the range $V_{m1} \rightarrow 0$, until the polarity of the stimulus  changes and consequently OVs  reverse their motion.  For $V=$ -1.4 V a SET transition to the LR$_1$ state takes place. The associated OVs distribution is shown in  panel II) (row labeled LR$_1$) with the color level in each of the L, C and R regions scketched in  panel III)(d),  proportional to the total OV density respectively. Notice that the  R layer is mostly depleted from OVs which accumulate in the C layer. In spite of  this accumulation, zone C still dominates the device resistance in the low resistance state LR$_1$. 
 Increasing further the intensity of the negative voltage, produces the electromigration of  OV from the C to the L layer, almost voiding of OVs the C layer and promoting   a second   RESET transition to the HR$_2$ state at a voltage $V=-$2 V (see also row labeled HR$_2$ in panel II)   and  panel III) (e)). The OVs density profile  for this case results slightly different from the previous one obtained for the HR$_1$ state, giving HR$_2$ $<$ HR$_1$, in fully agreement with  the experimental TWL-like HSL. The HR$_2$ state is mantained until the voltage  changes to possitive values and  the OVs electromigrate from the L to the C layer,  attaining the LR$_2$ state  at $V= 1.2$ V.

The agreement between the simulated and  the experimental TWL-like  HSL show in Figure \ref{f4} (c) is remarkable, denoting the predictive power of the VEOV model. 
For completness,   OV's density profiles  numerically obtained for other intermediate resistance states  are also displayed in different rows of panel II).

By changing to a non-symmetrical voltage protocol we  go from the TWL-like HSL to  squared HSL's,  emulating  the experimental response.
The CW squared HSL (blue) shown in  Figure \ref{f7} panel (I) was obtained for a voltage protocol $0\rightarrow$ $V_{m1}$=1.4 V$\rightarrow $ $-V_{m2}$ =2.1 V $\rightarrow 0$, whilts the  CCW HSL  (orange) corresponds to a voltage protocol $0\rightarrow$ $V_{m1}$=2.1 V$\rightarrow $ $-V_{m2} $=1.4 V $\rightarrow 0$.

\section {Discussion and conclusions}

The experiments and modeling described in the present work show that it is possible to selectively activate/deactivate two series memristive interfaces in order to obtain squared CW, CCW or a  TWL-like HSLs just by controlling the maximum ($V_{max} >$0) and minimum ($V_{min}<$0) applied voltage excursions. This behaviour is independent of the nature of the used electrodes: we found it in Pt/TaO$_{2-h}$/Ta$_2$O$_{5-x}$/TaO$_{2-y}$/Pt devices, but also in similar systems where the top Pt electrode was replaced by Au, as reported in Fig. S5 of the Suppl. Information. 

A key factor that allows the obtained  selective response is a forming process that changes the initial asymmetric Ta-oxide bilayer, grown at different oxygen pressures, to a quasi-symmetric TaO$_{2-h}$/Ta$_2$O$_{5-x}$/TaO$_{2-y}$ stack, excluding the Pt(Au)/TaO$_{2-h(y)}$ metal-metal interfaces from contributing to the memristive behavior. The finding of CCW and CW HSLs in a single quasi-symmetric device goes against the common wisdom that, for interface-related memristors, ``(a)symmetric systems give (a)symmetric electrical response". We propose, based on our thourough analysis and modeling, that the  physical origin of the obtained electrical behavior relies on OVs electromigration between bulk and interfacial zones of the Ta$_2$O$_{5-x}$ layer, sandwiched between metallic TaO$_{2-h(y)}$. From the combination of experimental results and simulations it can be stated that for CCW HSL the OVs dynamics is constrained to C and L zones, while for CW HSL the OV exchange is limited between C and R layers. For symmetric stimuli all L, C and R are involved and the memristive response becomes symmetric (TWL-like HSL). 

It can be concluded that OV transfer between one interface (i.e. R) and C zone does not start until the other interface (i.e. L) is almost completely drained of OV. This 2-steps process is the core of the observed behavior. It is worth to remark that this rich phenomenology was found for a system with a rather simple room temperature fabrication process that includes only one oxide -Ta-oxide- and one metal, suggesting that it might be possible to take advantage of these properties at low cost and with easy scability. 

Regarding possible applications, the control of the symmetry of the electric response allows optimizing, for example, ON-OFF ratios for RRAM memories; however, the results presented here have potentially more implications for the development of disruptive electronics such as neuromorphic computing or novel logic devices. The observation of multilevel resistance states indicates that our devices can mimic the (analogic) adaptable synaptic weight of biological synapsis. Importantly, the multilevel states were found for both CCW and CW HSLs, indicating that the synaptic weight could be either potentiated or depressed with electrical stimulus of the same polarity. This might have implications for novel, beyond von Neummann, devices \cite{zhang_2018}.

\section {Experimental section}

Ta-oxide  thin films were deposited on commercial platinized silicon by pulsed laser deposition at room temperature from a single phase Ta$_2$O$_5$ ceramic target. We used a 2-steps deposition process with oxygen pressures of 0.01 and 0.1mbar, resulting in 35nm and 15nm layers, respectively, as shown in the main text. The laser fluence was fixed at 1.5 J/cm$^2$. For spectroscopic measurements, we also deposited single TaO$_x$ layers at 0.01 and 0.1mbar of O$_2$, respectively. A Dual Beam Helios 650 was used to acquire scanning electron microscopy images and prepare FIB lamellas to observe cross-sections. High resolution scanning transmission electron microscopy was performed using a FEI Titan G2 microscope at 300 kV with probe corrector and in situ EELS spectrum acquisition with a Gatan Energy Filter Tridiem 866 ERS. The surface composition analysis obtained by X-Ray Photoelectron Spectroscopy (XPS) used a standard Al/Mg twin anode X-ray gun and a hemispherical electrostatic electron energy analyzer. The base chamber pressure was 10-9 mbar. Top Pt electrodes, ~20nm thick, were deposited by sputtering and microstructured by optical lithography. Top electrode sizes ranged between 6.4x10$^3$ and 78.5x10$^3$ $\mu$m$^2$. Electrical characterization was performed at room temperature with a Keithley 2612 source-meter connected to a Suss probe station. 

\section*{acknowledgements}
We acknowledge support from INN-CNEA, UNCuyo (06/C455), ANPCyT (PICT2014-1382, PICT2016-0867, PICT2017-1836, PICT 2017-0984), CONICET (PIP
11220150100653CO) and Univ. Buenos Aires (UBACyT 20020170100284BA). M.A. also acknowledges financial support of H2020-MSCA-RISE-2016 SPICOLOST Grant No. 734187 to perform TEM studies at LMA-INA, University of Zaragoza. S.B. thanks technical support from N. Cortes. 

\bibliography{references}

\end{document}